\documentclass[entropy,article,submit,moreauthors,pdftex,12pt,a4paper]{mdpi}
\lastpage{x}
\doinum{10.3390/------}
\pubvolume{xx}
\pubyear{2015}
\history{Received: xx / Accepted: xx / Published: xx}

\usepackage{graphicx}
\usepackage{bm}
\usepackage{amsmath, amsthm, amssymb}
\usepackage{eurosym}
\usepackage{tabularx}

\Title{A search for a spectral technique to solve nonlinear fractional differential equations}
\Author{Malgorzata Turalska $^{1,\dagger}$* and Bruce J. West $^{1,2,\dagger}$ }
\address{
$^{1}$ Department of Physics, Duke University, Durham, NC 27709, USA\\
$^{2}$ Information Science Directorate, US Army Research Office, Research Triangle Park, NC 27708, USA}
\contributed{$^\dagger$ These authors contributed equally to this work.}
\corres{mat51@phy.duke.edu, 9196602582}

\abstract{
A spectral decomposition method is used to obtain solutions to a class of nonlinear differential equations. We extend this approach to the analysis of the fractional form of these equations and demonstrate the method by applying it to the fractional Riccati equation, the fractional logistic equation and a fractional cubic equation. The solutions reduce to those of the ordinary nonlinear differential equations, when the order of the fractional derivative is $\alpha =1$. The exact analytic solutions to the fractional nonlinear differential equations are not known, so we evaluate how well the derived solutions satisfy the corresponding fractional dynamic
equations. In the three cases we find a small, apparently generic, systematic error that we are not able to fully interpret. }


\begin{document}

\section{Introduction}

Herein we propose a spectral method for solving fractional nonlinear rate equations of a certain kind. The method is not perturbative, but neither is it exact, since it gives rise to systematic deviations of the analytic solution from the numerical solution at intermediate times that reaches a maximum value of 2\%. On the one hand, the spectral method provides a remarkable good approximation to numerical calculation. On the other hand, the source of the small but systematic deviation from the numerical solution remains a mystery. This paper presents the approach in detail and introduces a new problem that requires explanation.

Despite the advances made into the understanding of complex nonlinear systems in the last half of the twentieth century, many physical phenomena failed to be described using the tools of ordinary calculus. Nonlocal distributed effects and memory effects observed in relaxation phenomena \cite{melo04}, living systems \cite{magin10,henry08}, wave propagation in porous materials \cite{garra11} have been more successfully modeled within the framework of the fractional calculus \cite{stiassnie79}. Fractional differential equations (FDE) have been adopted to explain these and other complex phenomena \cite{west14,west15a}. Since exact solutions to the majority of FDEs are not available, the search for appropriate analytical and numerical methods is a subject of ongoing research. Recently, a number
of approaches devoted to solving FDEs have been proposed. Examples include Adomian decomposition method \cite{abba2006}, homotopy perturbation method \cite{khan2011, amin2010}, the fractional sub-equation method. and the Haar wavelet method \cite{li2010}, to name but a few. However, the convergence region of solutions obtained with these algorithms is rather small.

It was hypothesized \cite{svenkeson15} that the spectral decomposition method can be extended to the analysis of a class of nonlinear fractional differential equations (NFDEs). Herein we demonstrate the method by applying it to the fractional Riccati equation (FRE), the fractional logistic equation (FLE) and a fractional cubic equation (FCE), where the fractional-order is in the range $0<\alpha \leq 1$. The solutions obtained are shown to have the correct short-time and long-time behaviors. Additionally, they reduce to the well known solutions of the ordinary nonlinear differential equations, when the order of the fractional derivative is $\alpha =1$. In the cases considered herein the exact analytic solution to the FNDE was not known previously, we evaluate how well the derived solutions satisfy the corresponding NFDE using numerical techniques. In all cases we find a very small, but systematic deviation of the analytic
from the numerical solutions that has eluded our best efforts to interpret. One possibility, of course, is that the numerical technique used to solve the NFDE is the culprit, since it was based on numerically solving linear fractional equations. But this remains to be investigated.

In Section \ref{spectral} we introduce the spectral decomposition of the solution to define the eigenvalue problem for integer-order linear and nonlinear rate equations, as well as NFDEs. In Section \ref{nonlinear} we obtain a series expansion over the spectrum of eigenvalues and eigenfunctions for the solution to three NFDEs, where the exponentials in the solutions to the integer-order equations, also obtained, are replaced with Mittag-Leffler functions (MLFs). Exact solutions to NFDEs are rare in the literature \cite{zhang10,west15}, so to test the validity of the analytic results we numerically evaluate how the solutions obtained satisfy
the appropriate NFDE. To our surprise the error function measuring this fit is not zero, but varies in time, increasing as $t^{2\alpha }$ at early times and decreasing as $t^{-\alpha }$ at late times, and reaching a maximum difference of less than a few percent at an intermediate time. This non-monotonic scaling difference is shown to occur with the solutions to the FRE, the FLE, as well as, the FCE, all with the same qualitative behavior in the error. In Section \ref{discussion} we draw some tentative conclusions including the speculation that this systematic deviation may be generic.

\section{Spectral decomposition\label{spectral}}

\subsection{Integer operator}

Let us begin by establishing the nomenclature used in the study of the nonlinear differential equations. Consider the one-dimensional first-order differential equation
\begin{equation}
\frac{d}{dt}X(t)=\mathcal{O}X(t),  \label{basic}
\end{equation}%
where $X(t)$ is the dynamic variable of interest and $\mathcal{O}$ is a generic operator acting on $X(t)$. Allowing Eq. (\ref{basic}) to describe any dynamical system of interest entails the formal solution
\begin{equation}
X(t)=e^{\mathcal{O}_{0}t}x_{0},  \label{basic_operator}
\end{equation}%
where $x_{0}\equiv X(0)$ defines the initial condition in the phase space for the dynamic variable and the operator $\mathcal{O}_{0}$ acts on the initial condition. The exponential operator is formally defined by the series expansion
\begin{equation}
e^{\mathcal{O}_{0}t}=\sum_{k=0}^{\infty}\frac{\left( \mathcal{O}_{0}t\right) ^{k}}{\Gamma \left( k+1\right) }  \label{exp_op}
\end{equation}%
so that the solution Eq.(\ref{basic_operator}) can be expressed as
\begin{equation}
X(t)=\sum_{k=0}^{\infty }\frac{\left( \mathcal{O}_{0}t\right) ^{k}}{\Gamma \left( k+1\right) }x_{0}=\sum_{k=0}^{\infty }\frac{t^{k}}{%
\Gamma \left( k+1\right) }\mathcal{O}_{0}{}^{k}x_{0}  \label{sol_exp}
\end{equation}%
where the operator $\mathcal{O}_{0}{}^{k}$\ acts solely on the initial condition. Note that for a linear equation with a constant rate $\lambda$ the operator is given by
\begin{equation}
\mathcal{O}_{0}{}^{k}x_{0}=\left( \lambda x_{0}\frac{\partial }{\partial x_{0}}\right) ^{k}x_{0}=\lambda ^{k}x_{0},
\label{linear eigenvalue}
\end{equation}%
which when inserted into Eq.(\ref{sol_exp}) and summing the series yields the exponential solution to the scalar rate equation
\begin{equation}
X(t)=e^{\lambda t}x_{0}.  \label{exp sol}
\end{equation}
It is apparent that Eq.(\ref{linear eigenvalue}) has a form suggestive of an eigenvalue equation and that the solution to the general integer-operator rate equation can be expressed as an eigenfunction expansion over the spectrum of eigenvalues%
\begin{equation}
X(t)=\overset{\infty }{\underset{k=0}{\sum }}C_{k}\phi _{k}\left(x_{0}\right) \chi _{k}\left( t\right) .  \label{eigensolution}
\end{equation}%
The quantity $\phi _{k}\left( x_{0}\right) \chi _{k}\left( t\right) $ is the eigenfunction, factored into a piece determined by the spectrum of eigenvalues $\left\{ \lambda _{k};k=0,1,2,\cdot \cdot \right\} $, a piece determined by the initial condition $x_{0}$, and the expansion coefficient $C_{k}$ determined by the dynamics and overall initial normalization. Inserting Eq.(\ref{eigensolution}) into (\ref{basic}), allows us to separate out the time-dependence of the components of the expansion%
\begin{equation}
\frac{d}{dt}\chi _{k}\left( t\right) =\mathcal{\lambda }_{k}\chi _{k}\left(t\right) \Rightarrow \chi _{k}\left( t\right) =e^{\lambda _{k}t}.  \label{eigenfunction}
\end{equation}%
Correspondingly, the eigenvalue equations are given by \begin{equation}
\mathcal{O}_{0}\phi _{k}\left( x_{0}\right) =\mathcal{\lambda }_{k}\phi_{k}\left( x_{0}\right)   \label{ev equation}
\end{equation}%
and the eigenvalues are determined by the form of the operator.

In the linear case just considered the operator is the same as before, so the equation for the eigenfunction is%
\begin{equation*}
\lambda x_{0}\frac{d\phi _{k}}{dx_{0}}=\lambda _{k}\phi _{k}
\end{equation*}%
with the solution
\begin{equation*}
\phi _{k}\left( x_{0}\right) =x_{0}^{\frac{\lambda _{k}}{\lambda }}.
\end{equation*}%
The linear eigenvalue spectrum is degenerate $\lambda _{k}=\lambda $ and the coefficients are determined from the initial condition to satisfy%
\begin{equation*}
\overset{\infty }{\underset{k=0}{\sum }}C_{k}=1.
\end{equation*}%
The resulting solution is, of course, given by Eq.(\ref{exp sol}).

\subsection{Non-integer (fractional) operator}

Now assume that this general form of a solution to a differential equation translates to the fractional calculus domain. Thus, we replace Eq.(\ref{basic}) with the fractional differential equation
\begin{equation}
\frac{d^{\alpha }}{dt^{\alpha }}X(t)=\mathcal{O}X(t),  \label{frac_basic}
\end{equation}%
where $0<\alpha \leq 1$. We assume the fractional derivative to be defined in the Caputo sense:
\begin{equation}
\frac{d^{\alpha }}{dt^{\alpha }}X(t)=\frac{1}{\Gamma \left( 1-\alpha \right)
}\int_{0}^{t}\frac{X^{\prime }(\tau )}{\left( t-\tau \right) ^{\alpha }}%
d\tau .  \label{caputo}
\end{equation}%
where $X^{\prime }(\tau )$ denotes the derivative of $X(\tau )$ with respect to its argument. Eq.(\ref{frac_basic}) can be solved analytically in terms of the MLF by employing the spectral decomposition introduced above in which case we have for the components of the eigenfunction%
\begin{equation}
\frac{d^{\alpha }}{dt^{\alpha }}\chi _{k}(t)=\lambda _{k}\chi _{k}\left(t\right) ,  \label{eigenfunction2}
\end{equation}%
to obtain the MLF evaluated over the spectrum of eigenvalues
\begin{equation}
\chi _{k}(t)=E_{\alpha }\left( \mathcal{\lambda }_{k}t^{\alpha }\right) .
\label{sol_frac}
\end{equation}%
The MLF is defined by the series \cite{west03,podlubny99}%
\begin{equation}
E_{\alpha }\left( z\right) =\underset{k=0}{\overset{\infty }{\sum }}\frac{%
z^{k}}{\Gamma \left( k\alpha +1\right) }.  \label{MLF}
\end{equation}%
Consequently, inserting the MLF into the expansion for the solution yields \cite{svenkeson15}%
\begin{equation}
X(t)=\overset{\infty }{\underset{k=0}{\sum }}C_{k}\phi _{k}\left(
x_{0}\right) E_{\alpha }\left( \mathcal{\lambda }_{k}t^{\alpha }\right)
\label{nonlinear expansion}
\end{equation}%
and the eigenvalues are determined by the operator in Eq.(\ref{ev equation}). The MLF reduces to an exponential function when $\alpha =1$, reducing the series expansion to the ordinary solution of Eq.(\ref{basic_operator}) in that case.

Note that we can adopt the same formal spectral decomposition employed for the integer-derivative case discussed above for the fractional-order dynamics considered here. But, before we explore the fractional case, let us examine an integer-order nonlinear dynamic equation.

\section{Riccati equation\label{nonlinear}}

We illustrate the spectral decomposition method, applied to a nonlinear rate equation, using the reduced form of the Riccati equation \cite{davis62}:
\begin{equation}
\frac{d}{dt}X(t)=1-X^{2}(t)=\mathcal{O}X(t).  \label{riccatiE}
\end{equation}%
The Riccati equation is put into the form of Eq.(\ref{basic}) by introducing the phase space operator $\mathcal{O}$ :%
\begin{equation}
\mathcal{O}=\left( 1-x^{2}\right) \frac{\partial }{\partial x},
\label{operator}
\end{equation}%
and the formal solution to the Riccati equation can be written as the eigenmode expansion
\begin{equation}
X(t)=\sum_{k=0}^{\infty }C_{k}\phi _{k}\left( x_{0}\right)
e^{\lambda _{k}t}.  \label{riccati_sol}
\end{equation}%
The dependence of the solution on the initial condition can be described by the eigenfunctions equation%
\begin{eqnarray}
\mathcal{O}_{0}\phi _{k}\left( x_{0}\right)  &\equiv &\left[ 1-x_{0}^{2}%
\right] \frac{d\phi _{k}\left( x_{0}\right) }{dx_{0}}  \notag \\
&=&\lambda _{k}\phi _{k}\left( x_{0}\right) .  \label{eigenfunction3}
\end{eqnarray}%
The solution to Eq.(\ref{eigenfunction3}) is given by%
\begin{equation*}
\ln \phi _{k}\left( x_{0}\right) =\frac{\lambda _{k}}{2}\ln \left[ \frac{%
1+x_{0}}{1-x_{0}}\right] ,
\end{equation*}%
where we need to determine the spectrum of eigenvalues $\left\{ \lambda_{k}\right\} .$

This last expression is inserted into Eq. (\ref{riccati_sol}) to obtain
\begin{equation}
X(t)=\sum_{k=0}^{\infty }C_{k}\left[ \frac{1+x_{0}}{1-x_{0}}\right] ^{\frac{%
\lambda _{k}}{2}}e^{\lambda _{k}t}.  \label{test sol}
\end{equation}%
It is straightforward to obtain the eigenvalues $\lambda _{k}=-2k$ by inserting Eq.(\ref{test sol}) into Eq.(\ref{riccatiE}) and equating coefficients of time dependent terms, as well as the coefficients for equal $k$ values%
\begin{equation*}
C_{k}=2\left( \frac{C_{1}}{2}\right) ^{k};k>0.
\end{equation*}%
The series solution is then%
\begin{equation*}
X(t)=2\sum_{k=0}^{\infty }\left( \frac{C_{1}}{2}\right) ^{k}\left( \frac{%
1-x_{0}}{1+x_{0}}\right) ^{k}e^{-2kt}.
\end{equation*}%
The final expansion coefficient is determined to be $C_{1}=-2$ from the initial condition, so that the solution to the nonlinear initial value problem becomes
\begin{equation}
X(t)=2\sum_{k=0}^{\infty }\left( \frac{x_{0}-1}{x_{0}+1}\right)
^{k}e^{-2kt}-1  \label{riccati_sol3}
\end{equation}

Additionally, for $x_{0}=0$, Eq.(\ref{riccati_sol3}) reduces to the well-known solution of Riccati equation \cite{davis62}
\begin{equation*}
X(t)=\tanh (t).
\end{equation*}%
For initial values $x_{0}\neq 0$ the solution to Eq. (\ref{riccatiE}) is obtained by summing the series in Eq.(\ref{riccati_sol3}) to obtain
\begin{equation}
X(t)=\frac{\tanh (t)+x_{0}}{x_{0}\tanh (t)+1}  \label{riccati_sol4}
\end{equation}%
also in agreement with the solution obtain by Davis \cite{davis62}.

Since we have an exact solution we can test the series expansion computationally. The error is defined as the difference between the time derivative of the series solution obtained using Eq.(\ref{riccati_sol3}) and the known exact solution given by Eq.(\ref{riccati_sol4}) inserted into the right hand side of the Riccati equation:
\begin{equation}
\Delta (t)=2\sum_{k=0}^{\infty }\left( \frac{x_{0}-1}{x_{0}+1}\right)
^{k}\left( -2k\right) e^{-2kt}-1+\left[ \frac{\tanh (t)+x_{0}}{x_{0}\tanh
(t)+1}\right] ^{2}.
\end{equation}%
The sum is estimated by its first 100 terms yielding a difference on the order of $10^{-15}$, which matches machine precision, affirming that the series given by Eq. $\left( \ref{riccati_sol3}\right) $ is a valid solution to the Riccati equation.

\section{Fractional Riccati equation}
\subsection{Spectral decomposition solution}

We now establish that a generalization of the spectral decomposition method leads to valid solutions for certain NFDEs. We start from the fractional form of the Riccati equation
\begin{equation}
\frac{d^{\alpha }}{dt^{\alpha }}X(t)=1-X^{2}(t)=\mathcal{O}X(t)
\label{riccati_frac}
\end{equation}%
where $0<\alpha <1$. If the fractional derivative is of the Caputo type, Eq.(\ref{riccati_frac}) has the operator given by Eq.(\ref{operator}). Consequently, the solution to the FRE is of the form
\begin{equation}
X(t)=sum_{k=0}^{\infty }C_{k}\phi _{k}\left( x_{0}\right) E_{\alpha}\left( \lambda _{k}t\right) ,  \label{frac_sol1}
\end{equation}%
and the eigenvalue problem is the same as for the integer-order Riccati equation, that is, given by Eq.(\ref{eigenfunction3}). Consequently, we have the same spectrum of eigenvalues obtained in the integer-order RE, obtained at early times using the stretched form of the MLF. In the same way the expansion coefficients are the same as previously obtained in order to satisfy the initial condition and we obtain for the solution to the FRE:
\begin{equation}
X(t)=2\sum_{k=0}^{\infty }\left( \frac{x_{0}-1}{x_{0}+1}\right)
^{k}E_{\alpha }\left( -2kt^{\alpha }\right) -1.  \label{frac_sol2}
\end{equation}

We note that the solution obtained to the FRE does not coincide with that obtained by Zhang \textit{et al}. \cite{zhang10}. However we point out that these authors use Jumarie's modified form of the Reimann-Liouville fractional derivative \cite{jumarie06}, which explicitly satisfies the Leibniz condition:%
\begin{equation*}
\frac{d^{\alpha }}{dt^{\alpha }}\left[ f(t)g(t)\right] =g(t)\frac{d^{\alpha
}f(t)}{dt^{\alpha }}+f(t)\frac{d^{\alpha }g(t)}{dt^{\alpha }}
\end{equation*}%
for the derivative of the product of two analytic functions $f(t)$ and $g(t)$ . It is evident that the Caputo fractional derivative does not satisfy the Leibniz condition \cite{podlubny99,west03} and consequently, although the starting dynamic equations look the same, that because of the restrictions on the fractional derivatives, the problem solved by Zhang et al. \cite{zhang10} is different from the FRE considered here.

\begin{figure}[t]
\includegraphics[scale=0.90]{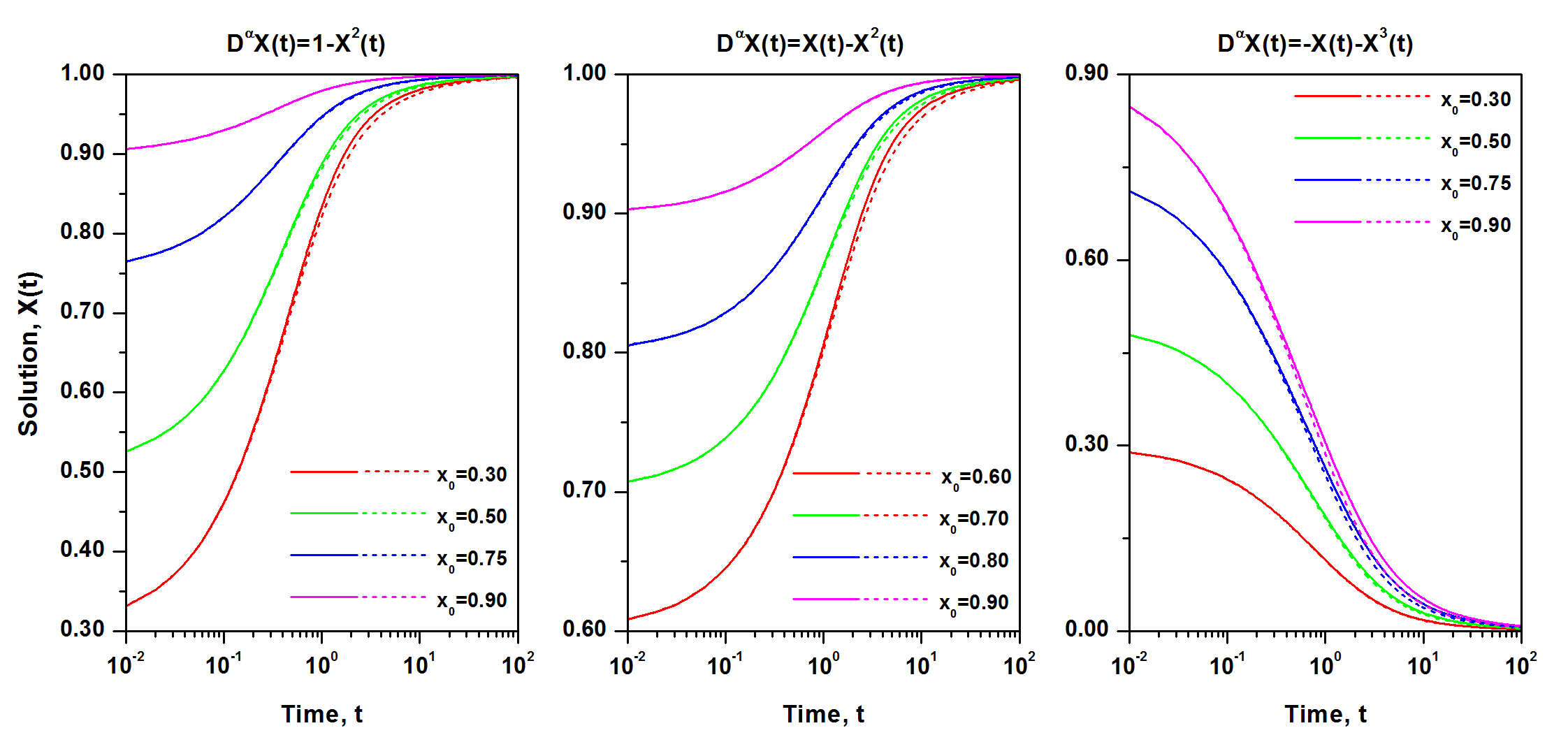}
\caption{ Solutions to the fractional differential equations. \textit{(left)} Fractional Riccati differential equation, as defined by Eq. \protect\ref{frac_sol2}. \textit{(middle)} Fractional logistic equation, as defined by Eq. \protect\ref{log_sol2}. The growth rate is $\protect\lambda =1.00.$ \textit{(right)} \ The cubic fractional equation, as defined by Eq. \protect\ref{cubic_sol}, with $a=b=1.00.$ The order of the fractional derivative in all cases is $\protect \alpha =0.75$. Continuous lines correspond to solutions obtained with the spectral decomposition method, dashed lines correspond to numerical integration of given equations. The integration step is $h=10^{-3}.$ }
\label{fig_solution}
\end{figure}

We see in this figure that the analytic solution does extremely well in following the numerical solution to the FRE, but it is not exact. The numerical solutions for various initial conditions, obtained using Eq.(\ref{frac_sol2}), where the infinite sum is approximated by its first 100 elements, are demonstrated in the first \ panel of the tryptic in Figure \ref{fig_solution}. One can easily verify that the proposed solution satisfies the initial condition and has the correct long time behavior. But since there are no known exact solutions to the initial value problem for the NRE, we use the numerical test introduced previously to test the veracity of Eq.(\ref{frac_sol2}).

\subsection{Testing the solution}

Since the exact analytic solution to the FRE Eq. (\ref{riccati_frac}) is not known, one possible approach to check the validity of Eq.(\ref{frac_sol2}) is to check that this solution satisfies the fractional differential equation in question. Applying the Caputo definition of fractional derivative to the analytic solution to the RME yields
\begin{equation}
LHS=\frac{d^{\alpha }}{dt^{\alpha }}X(t)=2\sum_{k=0}^{\infty }\left( \frac{%
x_{0}-1}{x_{0}+1}\right) ^{k}(-2k)E_{\alpha }\left( -2kt^{\alpha }\right) ,
\label{LHS}
\end{equation}%
while the right hand side of the FRE is
\begin{equation}
RHS=1-X^{2}(t)=1-\left( 2\sum_{k=0}^{\infty }\left( \frac{x_{0}-1}{x_{0}+1}%
\right) ^{k}E_{\alpha }\left( -2kt^{\alpha }\right) -1\right) ^{2}.
\label{RHS}
\end{equation}%
Again the potential error is defined in terms of the difference
\begin{equation}
\Delta (t)=RHS-LHS,  \label{diff1}
\end{equation}

\begin{figure}[t]
\includegraphics[scale=0.90]{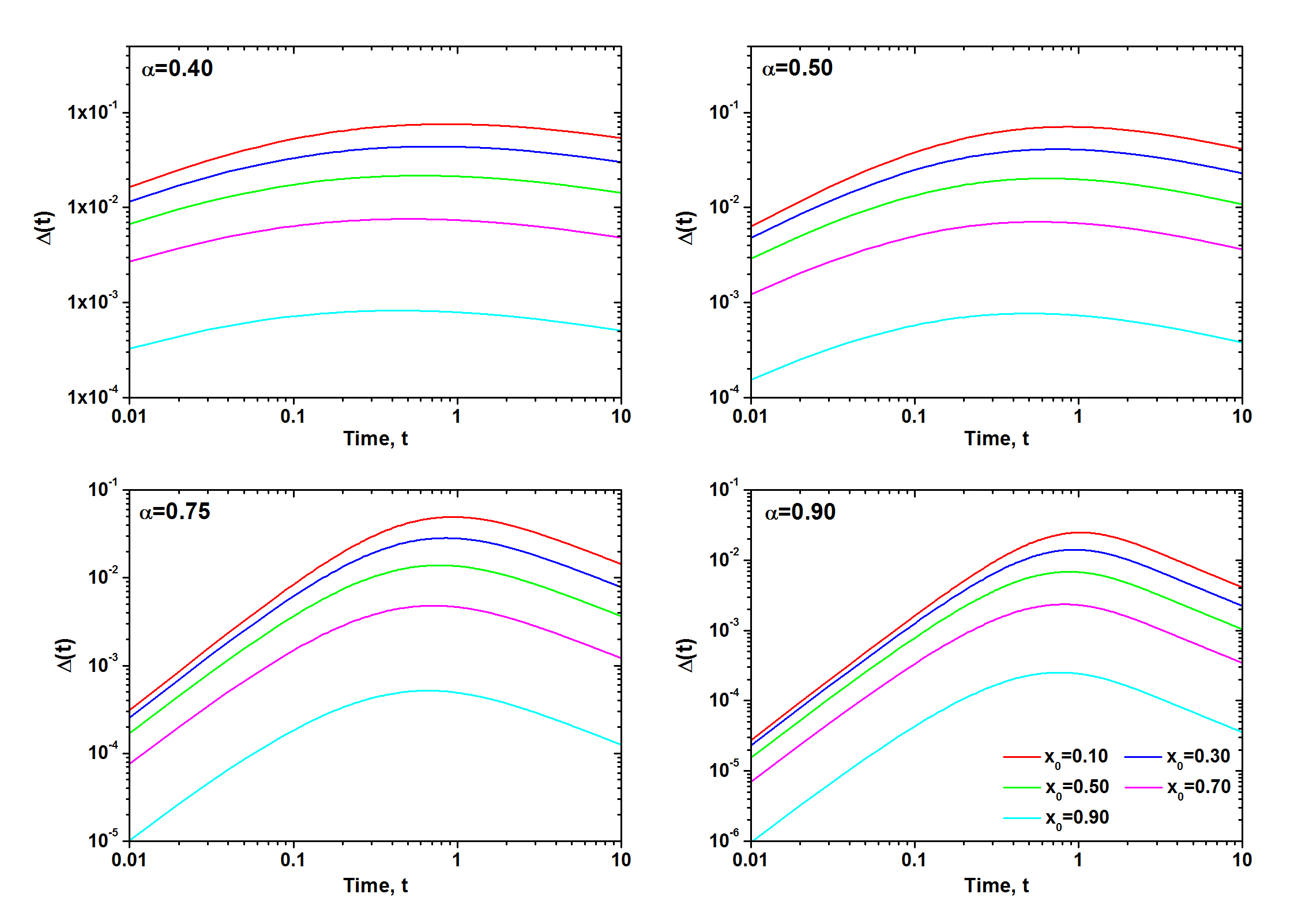}
\caption{ The difference $\Delta (t)$ between the RHS\ and LHS of the fractional Riccati equation, when the
solution is assumed to be Eq. \protect\ref{frac_sol2}. Consecutive panels correspond to an increasing order of the fractional derivative $\protect \alpha $. Color lines correspond to range of initial conditions, as denoted
by the legend in lower right plot. }
\label{fig_difference}
\end{figure}

which for an exact solution should be zero. This error variable is estimated numerically, and its behavior is plotted in Figure \ref{fig_difference} for four values of $\alpha $. It is apparent that the difference variable $\Delta (t)$ departs from zero at early times, gradually increasing, reaching a maximum value at $t\approx 1$, and finally decreases at long times. The behavior of $\Delta (t)$ and its maximum value $\Delta_{\max }<$ a few percent, clearly depends on the order of the fractional derivative, $\alpha ,$ and on the initial condition $x_{0}$. The precise form of power-law scaling of $\Delta (t)$, present at short and long times, is obtained adopting an approximation to the MLF. At early times, the series
defining the MLF (Eq.\ \ref{MLF}) reduces to a stretched exponential function
\begin{equation}
E_{\alpha }^{0}=\lim_{t\rightarrow 0}E_{\alpha }\left( -\lambda t^{\alpha
}\right) =1-\frac{\lambda t^{\alpha }}{\Gamma \left( 1+\alpha \right) }%
+...=\exp \left[ -\frac{\lambda t^{\alpha }}{\Gamma \left( 1+\alpha \right) }%
\right] ,  \label{MLFearly}
\end{equation}%
while at late times the MLF has inverse-power law behavior
\begin{equation}
E_{\alpha }^{\infty }=\lim_{t\rightarrow \infty }E_{\alpha }\left( -\lambda
t^{\alpha }\right) =\frac{t^{-\alpha }}{\lambda \Gamma \left( 1-\alpha
\right) }.  \label{MLFlong}
\end{equation}%
Thus, at early times the difference $\Delta (t)$ scales as $t^{2\alpha }$ and after some algebra this difference is determined to be
\begin{equation}
\Delta ^{0}(t)=\lim_{t\rightarrow 0}\Delta (t)=\frac{t^{2\alpha }}{\Gamma
^{2}\left( 1+\alpha \right) }\left( 1-x_{0}^{2}\right) ^{2},
\label{diffSHORT}
\end{equation}%
while the long-time behavior of the difference is determined by direct calculation to scale as $t^{-\alpha }:$
\begin{equation}
\Delta ^{\infty }(t)=\lim_{t\rightarrow \infty }\Delta (t)=\frac{t^{-\alpha }%
}{\Gamma \left( 1-\alpha \right) }\left[ 2\log \left( \frac{x_{0}+1}{2}%
\right) +x_{0}+1-\frac{\log ^{2}\left( \frac{x_{0}+1}{2}\right) }{\Gamma
\left( 1-\alpha \right) }t^{-\alpha }\right]  \label{diffLONG}
\end{equation}

\begin{figure}[t]
\includegraphics{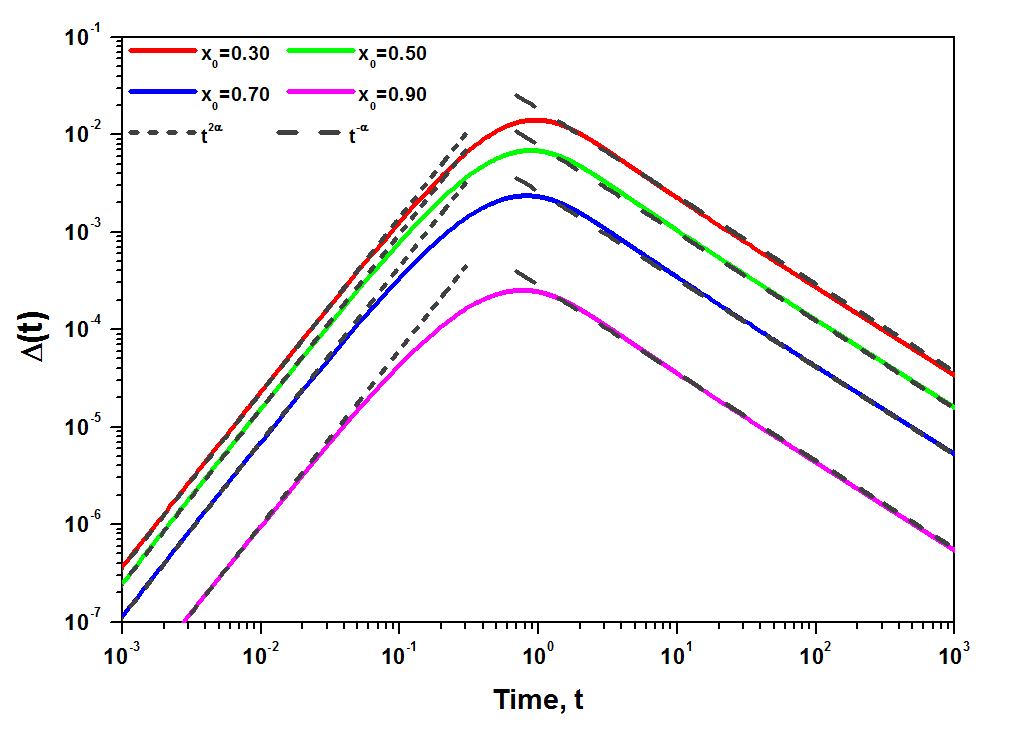}
\caption{ Short and long time power law behavior of the error function $\Delta (t)$. Both regimes are approximated
by equations \protect\ref{diffSHORT} and \protect\ref{diffLONG}, respectively. The order of the fractional derivative $\protect\alpha =0.90$. }
\label{fig_difference_scaling}
\end{figure}

Figure \ref{fig_difference_scaling} illustrates the validity of both approximations to the difference $\Delta (t)$ for selected orders of the fractional derivative. It is evident that the deviation of the solution obtained by spectral decomposition deviates from the exact solution in systematic ways at both late and early times. The maximum difference for $\alpha =0.9$ is shown in this figure to be below $2\%$, but the explanation as to why the deviation behaves the way it does remains elusive. Therefore we examine the solutions obtained by applying the technique to two other well known nonlinear initial value problems extended to the fractional domain.

\section{Fractional logistic equation}
\subsection{Spectral decomposition solution}

Next we examine the fractional logistic equation (FLE)
\begin{equation}
\frac{d^{\alpha }}{dt^{\alpha }}X(t)=\lambda ^{\alpha }X(t)\left(
1-X(t)\right)  \label{frac_log}
\end{equation}%
where $X(t)$ is the normalized population and $\lambda $ is the population growth rate. As before we define a phase space operator
\begin{equation}
\mathcal{O}=\lambda ^{\alpha }x\left( 1-x\right) \frac{\partial }{\partial x}
\label{log_oper}
\end{equation}%
and write the formal solution as Eq.(\ref{frac_sol1}). The eigenvalue equation is slightly different from that for the FRE%
\begin{eqnarray}
\mathcal{O}_{0}\phi _{k}\left( x_{0}\right)  &=&\lambda ^{\alpha
}x_{0}\left( 1-x_{0}\right) \frac{d\phi _{k}\left( x_{0}\right) }{dx_{0}}
\notag \\
&=&\lambda _{k}\phi _{k}\left( x_{0}\right)   \label{new eigenfunction}
\end{eqnarray}%
After some algebra we obtain for the eigenfunctions%
\begin{equation*}
\phi _{k}\left( x_{0}\right) =\left( \frac{x_{0}}{1-x_{0}}\right) ^{\frac{%
\lambda _{k}}{\lambda ^{\alpha }}}
\end{equation*}%
and after some algebra, using the spectral decomposition method, we arrive at the eigenvalues $\lambda _{k}=-k\lambda ^{\alpha }$, the expansion coefficients $C_{k}=C_{1}^{k}$ and from the initial condition $C_{1}=-1.$ Inserting these quantities into the formal solution Eq.(\ref{frac_sol1}) yields
\begin{equation}
X(t)=\sum_{k=0}^{\infty }\left( \frac{x_{0}-1}{x_{0}}\right) ^{k}E_{\alpha
}\left( -k\lambda ^{\alpha }t^{\alpha }\right) .  \label{log_sol2}
\end{equation}%
where again $x_{0}$ is the initial condition$.$The numerical solutions for various initial conditions, obtained using Eq.(\ref{log_sol2}), where the infinite sum is approximated by its first 100 elements, are demonstrated in the middle\ panel of the tryptic in Figure \ref{fig_solution}. As you might expect the solutions are qualitatively the same as in the first panel.

When $\alpha =1,$ Eq.(\ref{log_sol2}) reduces to the well-known solution to the logistic equation
\begin{equation}
X(t)=\frac{x_{0}}{x_{0}+\left( 1-x_{0}\right) e^{-\lambda t}},
\label{log_classical}
\end{equation}%
yielding sigmoidal population growth from the initial value $X(0)=x_{0}$ to the saturation level $X(\infty )=1$.

Note that even though the FRE and the FLE have quadratic nonlinearities, the spectra in the two case are different. The former increasing as $2k$ and the latter as $k$, which also determines the difference in the normalization parameter $C_{1}$ in the two cases.

The solution, given by Eq.(\ref{log_sol2}), is identical to the one obtained by one of the authors \cite{west15} using the Carleman embedding technique. The Carleman embedding approach has successfully determined the solution to many nonlinear differential equations as presented and discussed by Kowalski and Steeb \cite{kowalski91}. The technique was applied to a FNDE in \cite{west15} to obtain an infinite-order hierarchy of fractional moment equations, using Laplace transforms and solved using a matrix diagonalization method. A modest deviation of the analytic solution from the numerical integration of the FLE was noted in this earlier analysis.

\begin{figure}[t]
\includegraphics[scale=0.90]{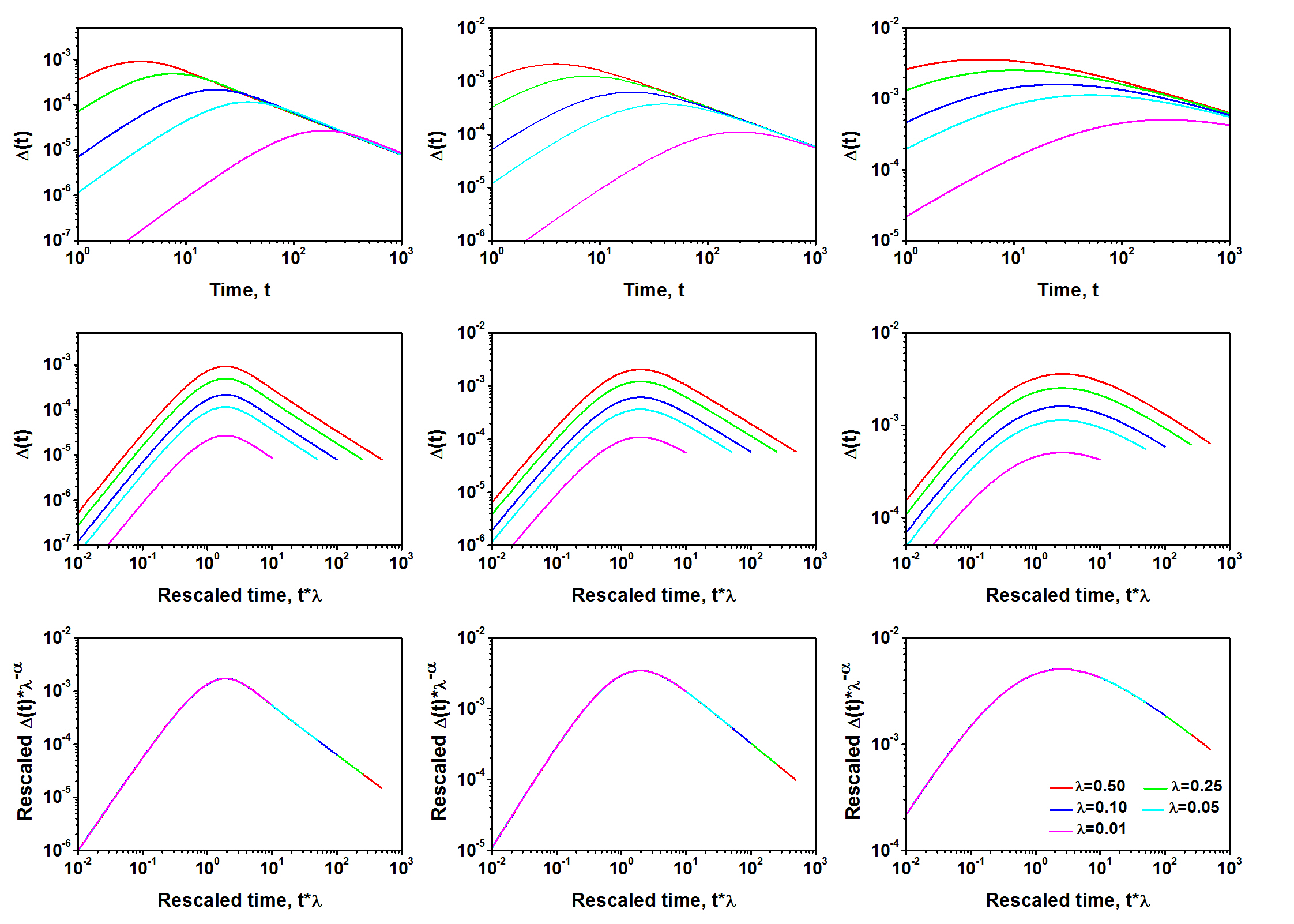}
\caption{ The difference $\Delta (t)$ between the RHS\ and LHS of the fractional logistic equation, when the solution is assumed to be Eq. \protect\ref{log_sol2}. Columns correspond to decreasing values of $\protect\alpha $: $\protect\alpha =0.90$ \textit{(left)}, $\protect\alpha =0.75$ \textit{(middle)} and $\protect\alpha =0.50$ \textit{(right)}. The initial value $x_{0}=0.75$ in all cases. Top row demonstrates the error function $\Delta (t)$ for a range of growth rate $\protect\lambda $ values. Middle row demonstrates the effect of rescaling time, while bottom row achieves complete overlap of $\Delta (t)$ curves generated for different $\protect\lambda $ through rescaling $y-$axis variable.}
\label{fig_log_difference_k}
\end{figure}

\subsection{Testing the solution}

Here again we check if the solution given by Eq.(\ref{log_sol2}) satisfies the FLE by examining the difference equation. The left hand side of Eq.(\ref{frac_log}) is
\begin{equation}
LHS=\frac{d^{\alpha }}{dt^{\alpha }}X(t)=\sum_{k=0}^{\infty }\left( \frac{%
x_{0}-1}{x_{0}}\right) ^{k}\left( -k\lambda ^{\alpha }\right) E_{\alpha
}\left( -k\lambda ^{\alpha }t^{\alpha }\right) ,  \label{LHS_log}
\end{equation}%
while the right hand side is
\begin{eqnarray}
RHS &=&\lambda ^{\alpha }X(t)\left( 1-X(t)\right)  \label{RHS_log} \\
&=&\lambda ^{\alpha }\sum_{k=0}^{\infty }\left( \frac{x_{0}-1}{x_{0}}\right)
^{k}E_{\alpha }\left( -k\lambda ^{\alpha }t^{\alpha }\right) \left(
1-\sum_{k^{\prime }=0}^{\infty }\left( \frac{x_{0}-1}{x_{0}}\right)
^{k^{\prime }}E_{\alpha }\left( -k^{\prime }\lambda ^{\alpha }t^{\alpha
}\right) \right) .
\end{eqnarray}%
The difference or potential error is defined as before by Eq.(\ref{diff1}).

Figure \ref{fig_log_difference_k} demonstrates the behavior of $\Delta (t)$ for selected values of the order of the fractional derivative $\alpha ,$ the initial condition $x_{0}$ and the unperturbed growth rate $\lambda $. We observe the same dependence of the difference $\Delta (t)$\ on time as in the case of the FRE. Note again that the the difference is not monotonic, but has a maximum value of less than 1\%, depending on $\alpha $ and $x_{0}$. The location of maximum in time is a function of $\lambda $. Rescaling both the $x-$axis and $y-$axis as follows
\begin{eqnarray}
t &\rightarrow &t\lambda  \label{rescaling} \\
\Delta (t) &\rightarrow &\Delta (t)\lambda ^{-\alpha }
\end{eqnarray}%
normalizes the $\Delta (t)$ curves, making them independent of the growth rate, as shown by the bottom row of curves in the figure.

Adopting the short-time and long-time approximations to the MLF we find that at early times the difference $\Delta (t)$ scales as $t^{2\alpha }:$
\begin{equation}
\Delta ^{0}(t)=\lim_{t\rightarrow 0}\Delta (t)=\frac{t^{2\alpha }}{\Gamma
^{2}\left( 1+\alpha \right) }x_{0}^{2}\left( x_{0}-1\right) ^{2},
\label{log_diffSHORT}
\end{equation}%
while the long time behavior scales as $t^{-\alpha }:$
\begin{equation}
\Delta ^{\infty }(t)=\lim_{t\rightarrow \infty }\Delta (t)=\frac{t^{-\alpha }%
}{\Gamma \left( 1-\alpha \right) }\left[ \left( x_{0}-1-\log x_{0}\right) -%
\frac{\log ^{2}x_{0}}{\Gamma \left( 1-\alpha \right) }t^{-\alpha }\right] .
\label{log_diffLONG}
\end{equation}

\begin{figure}[t]
\includegraphics[scale=0.90]{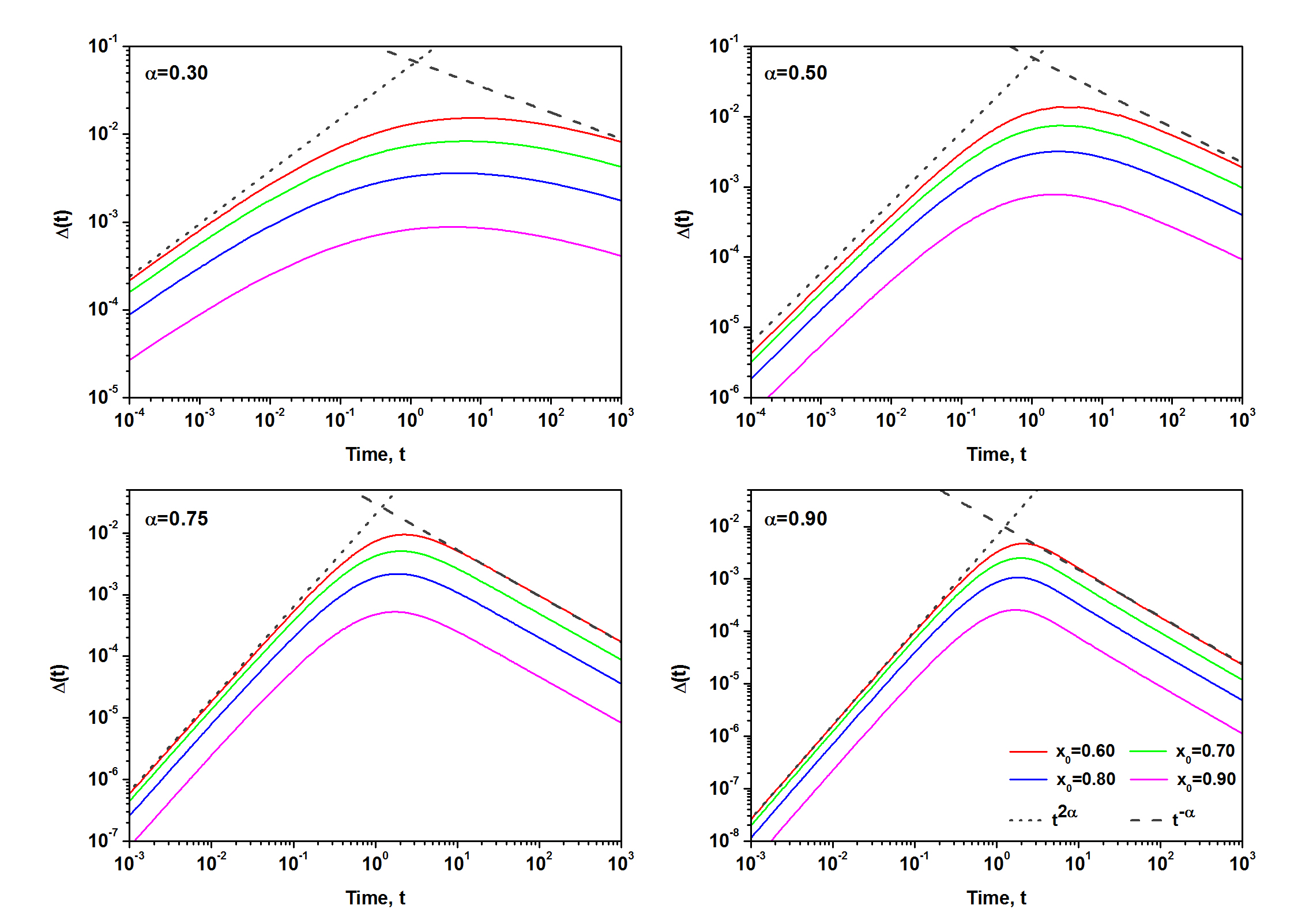}
\caption{ The difference $\Delta (t)$ between the RHS\ and LHS of the FLE. Consecutive panels correspond to an
increasing order of the fractional derivative $\protect\alpha $. Color lines correspond to a range of initial conditions, as denoted by the legend in lower right plot. The growth rate in all cases is $\protect\lambda =1.$ }
\label{fig_log_difference}
\end{figure}

The accuracy of the algebraically determined scaling is numerically demonstrated in Figure \ref{fig_log_difference}. Here again, the deviation of the solution determined using the spectral decomposition technique to solve the FLE from the exact numerical solution, is non-monotonic and never exceeds $1\%.$ Thus, the worst case using this technique is better than the
best obtained using many approximation techniques.

\section{Cubic fractional differential equation}
\subsection{Spectral decomposition solution}

The final equation that we consider is a fractional differential equation with a cubic term (FCE)
\begin{equation}
\frac{d^{\alpha }}{dt^{\alpha }}X(t)=-aX(t)-bX^{3}(t)=\mathcal{O}X(t).
\label{frac_cubic}
\end{equation}%
The formal solution given by the spectral decomposition requires the solution to the eigenvalue equation%
\begin{eqnarray}
\mathcal{O}_{0}\phi _{k}\left( x_{0}\right) &=&-x_{0}\left[ a+bx_{0}^{2}%
\right] \frac{d\phi _{k}\left( x_{0}\right) }{dx_{0}}  \notag \\
&=&\lambda _{k}\phi _{k}\left( x_{0}\right) .  \label{eigenvalue4}
\end{eqnarray}%
The algebra providing the solution to the eigenvalue problem will be presented elsewhere. Here we record that the solution is given by sum over the eigenvalue spectrum
\begin{equation}
X(t)=\sum_{k=0}^{\infty }\frac{\left( 2k-1\right) !!}{\left( 2k\right) !!}%
\left( \frac{\frac{b}{a}x_{0}^{2}}{\frac{b}{a}x_{0}^{2}+1}\right) ^{k}\frac{%
x_{0}}{\sqrt{\frac{b}{a}x_{0}^{2}+1}}E_{\alpha }\left( -\left( 2k+1\right)
at^{\alpha }\right) ,  \label{cubic_sol}
\end{equation}%
where we have adopted the notation of double factorial $\left( 2k\right)!!=2k\left( 2k-2\right) \left( 2k-4\right) \cdot \cdot \cdot.$ The solutions obtained using Eq. $\left( \ref{cubic_sol}\right) $ are presented on the right panel of Fig. \ref{fig_solution}.

For the integer-value case $\alpha =1$, the MLF can again be replaced by an exponential in Eq.(\ref{cubic_sol}) and the series summed to yield the exact integer-value solution%
\begin{equation}
X(t)=\frac{x_{0}e^{-at}}{\sqrt{1+\frac{b}{a}x_{0}^{2}\left(
1-e^{-2at}\right) }}.  \label{exact integer}
\end{equation}

\subsection{Testing the solution}

\begin{figure}[t]
\includegraphics[scale=0.90]{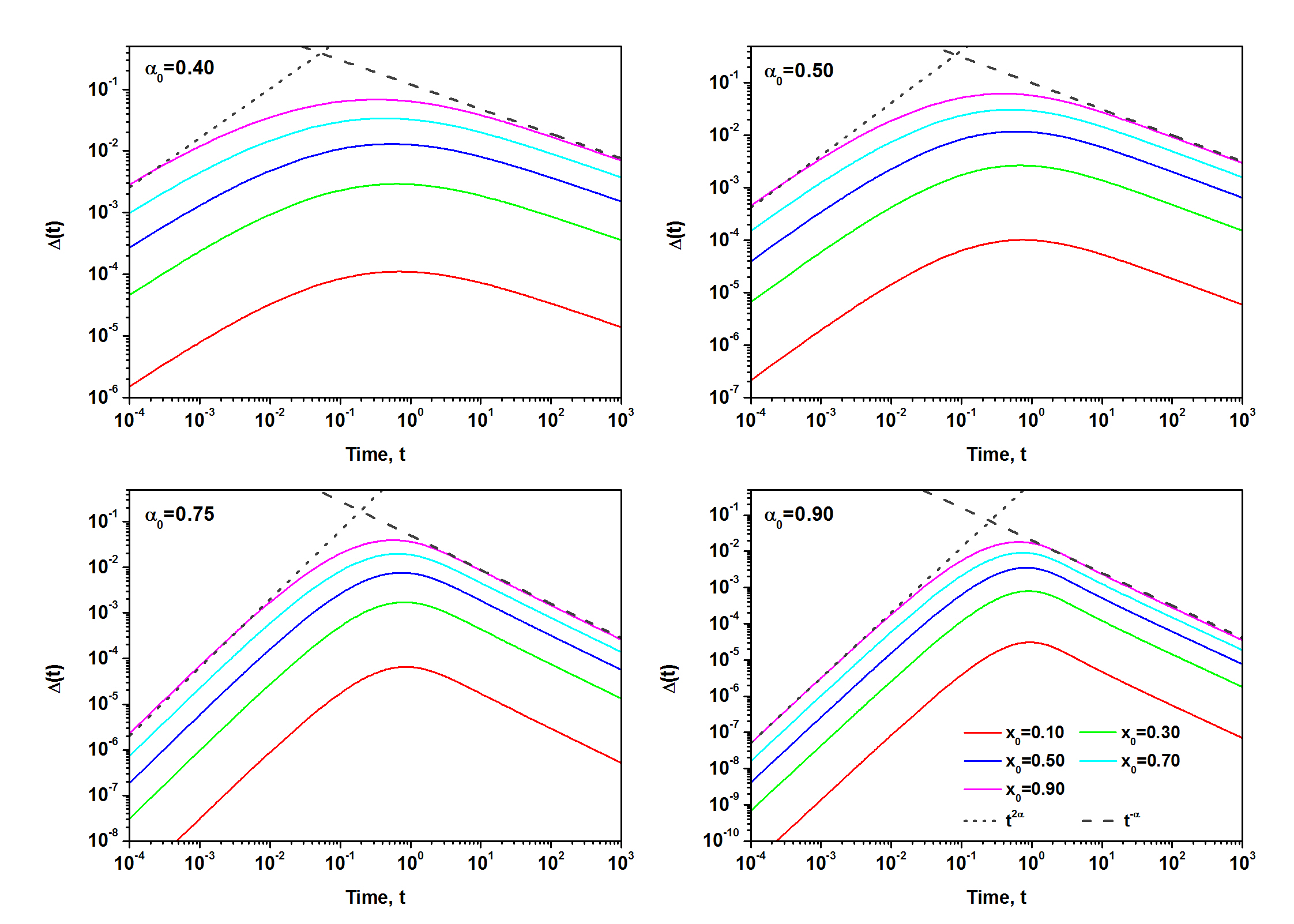}
\caption{ The difference $\Delta (t)$ between the RHS\ and LHS of the FCE. Consecutive panels correspond to an
increasing order of the fractional derivative $\protect\alpha $. Color lines correspond to range of initial conditions, as denoted by the legend in lower right plot. }
\label{fig_cubic_difference}
\end{figure}

Once again we check how well this solution satisfies the NFDE in question by taking the difference between the two sides of the FCE. The difference or potential error function $\Delta (t)$ is illustrated on Figure \ref{fig_cubic_difference} for $a=b=1$. The short-time behavior of $\Delta (t)$ is
\begin{equation}
\Delta ^{0}(t)=\lim_{t\rightarrow 0}\Delta (t)=\frac{t^{2\alpha }}{\Gamma
^{2}\left( 1+\alpha \right) }x_{0}^{3}\left( x_{0}-1\right) ^{2}\left[ 3-%
\frac{t^{\alpha }}{\Gamma \left( 1+\alpha \right) }\left( x_{0}-1\right) ^{3}%
\right] ,
\end{equation}%
while the long-time behavior is
\begin{equation}
\Delta ^{\infty }(t)=\lim_{t\rightarrow \infty }\Delta (t)=C_{1}\left(
\alpha ,x_{0}\right) \frac{t^{-\alpha }}{\Gamma \left( 1-\alpha \right) }%
+C_{2}\left( \alpha ,x_{0}\right) \frac{t^{-2\alpha }}{\Gamma ^{2}\left(
1-\alpha \right) }+C_{3}\left( \alpha ,x_{0}\right) \frac{t^{-3\alpha }}{%
\Gamma ^{3}\left( 1-\alpha \right) }
\end{equation}%
where $C_{1},C_{2}$ and $C_{3}$ are coefficients depending on $\alpha $ and $x_{0},$ and there is no need to record their values here.

Here again the deviation of the two sides of the FCE using the spectral decomposition analytic solution and the exact numerical calculation of the fractional derivative results in scaling as depicted in Figure \ref{fig_cubic_difference}. At early times the deviation increases as $t^{2\alpha }$ whereas at late times the deviation decreases as an inverse power law in time $t^{-\alpha }$, reaching a maximum at an intermediate time that does not exceed a few percent.

\section{Numerical solutions}

Since the analytic solutions to ordinary nonlinear differential equations are notoriously difficult to find, one turns to numerical methods for assistance. Due to the nonlocal property of the fractional derivatives, known numerical methods are not easily transferable to the fractional calculus. Herein we adopted the Adams-Bashforth-Moulton predictor-corrector
technique, developed by Diethelm e\textit{t al}. \cite{diethelm2002,diethelm2005} to investigate the solutions to fractional Riccati equation, fractional logistic equation and cubic fractional equation in purely numerical fashion through the numerical integration of the listed equations \ The main motivation is to compare the numerical solutions with the ones
obtained through spectral decomposition method.

\begin{figure}[t]
\includegraphics[scale=0.90]{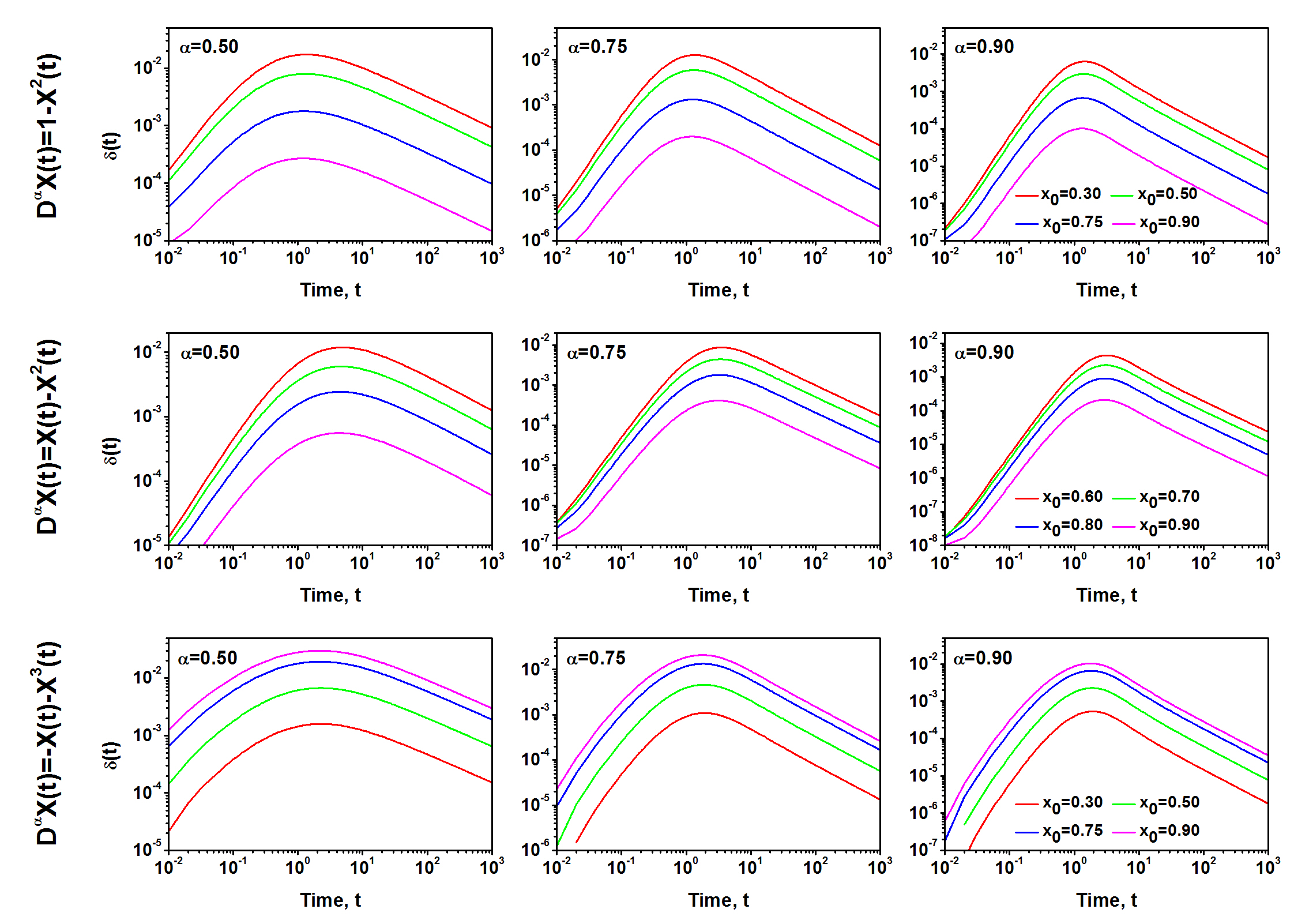}
\caption{ The difference $\protect\delta (t)$ between the numeric integration of fractional differential equations
and the solutions obtained with spectral decomposition method. }
\label{fig_numerical}
\end{figure}

Solutions for selected values of $\alpha $ and $x_{0}$ are presented on Fig. \ref{fig_solution}, together with the solutions obtained with the spectral decomposition. The overlap of both curves in all cases is extremely good. To better visualize how close the spectral decomposition solution $X_{SP}(t)$\ is to the numerical solution $X_{NUM}(t)$, on Figure \ref{fig_numerical} we plot the difference
\begin{equation}
\delta (t)=X_{NUM}(t)-X_{SP}(t).
\end{equation}%
In all cases the difference $\delta (t)$\ is smaller than $1\%$ and is characterized by an increase in short time scale, maximum value at intermediate times and and a decrease at large times, where the difference scales as $t^{-\alpha }$. This result, taken together with the behavior of $\Delta (t)$\ demonstrates that the spectral decomposition method provides
analytic solutions that are very close to the true solutions to the fractional nonlinear differential equations.

\section{Discussion\label{discussion}}

It has been demonstrated that the spectral decomposition solution to fractional nonlinear dynamic problems with quadratic and cubic nonlinearities systematically deviate from the evaluation of the fractional derivative in a common way. The systematic deviation is not monotonic, but for both quadratic and cubic nonlinearity grows as the same power law at
early time, decays as the same inverse power law at \ late times, and has a maximum deviation of the analytic solution from the numerical evaluation of the fractional derivative of a few percent at some intermediate time. Only the size of the coefficients change with the parameter values and the type of nonlinearity, but not the functional form in time. It appears that the scaling form of the deviation is a consequence of the MLF in the spectral decomposition, which changes as $t^{\alpha }$ at early times and as $t^{-\alpha }$ at late times.

Recall that the FLE was the only other case with which we could compare the solution obtained here with that obtained using a different method and they turned out to be identical. However, the hierarchy of equations solved using the matrix method to solve the FLE \cite{west15}, implicitly assumed a version of the Leibniz condition. The correspondence of that earlier solution and that obtained using the spectral decomposition method suggests that the deviation from the numerical solution, obtained using the latter technique is related to the Leibniz condition. Unfortunately we have not been able to identify precisely how this comes about and this remains a speculation.

\acknowledgments{Acknowledgments}
The authors would like to thank the U.S. Army Research Office for supporting this research.

\authorcontributions{Author Contributions}
MT and BJW conceived and designed the experiments, MT performed the experiments and analyzed the data; MT and BJW wrote the paper. 

\conflictofinterests{Conflicts of Interest}
The authors declare no conflict of interest.

\bibliographystyle{mdpi}

\begin{thebibliography}{99}

\bibitem{melo04}
Reyes-Melo, M.E.; Martinez-Vega, J.J.; Guerrero-Salazar, G.A.; Ortiz-Mendez, U. Modeling of relaxation phenomena in organic dielectric materials. Application of differential and integral operators of fractional order. \textit{J Optoelectronics and Adv Mat.} \textbf{2004}, \emph{6}, 1037-1043.

\bibitem{magin10}
Margin, R.L. Fractional calculus models of complex dynamics in biological tissues. \textit{Comps \& Math with Apps.} \textbf{2010}, \emph{59}, 1586-1593. doi: 10.1016/j.camwa.2009.08.039.

\bibitem{henry08}
Henry, B.I.; Langlands, T.A.M.; Wearne, S.L. Fractional cable models for spiny neuronal dendrites. \textit{Phys Rev Lett.} \textbf{2008}, \emph{100}, 128103. doi: http://dx.doi.org/10.1103/PhysRevLett.100.128103.

\bibitem{garra11}
Garra, R. Fractional-calculus model for temperature and pressure waves in fluid-saturated porous rocks. \textit{Phys Rev E.} \textbf{2011}, \emph{84}, 036605. doi: http://dx.doi.org/10.1103/PhysRevE.84.036605.

\bibitem{stiassnie79}
Stiassnie, M. On the application of fractional calculus for the formulation of viscoelastic models. \textit{Appl Math Modeling} \textbf{1979}, \emph{3}, 300-302. doi: doi:$10.1016/S0307-904X(79)80063-3$.

\bibitem{west14}
West, B.J. Colloquium: Fractional calculus view of complexity: A tutorial. \emph{Rev Mod Phys.} \textbf{2014}, \emph{86}, 1169. doi: http://dx.doi.org/10.1103/RevModPhys.86.1169.

\bibitem{west15a}
West, B.J. \emph{Fractional Calculus View of Complexity. Tomorrow's Science}, 1st ed.; CRC Pub.: Boca Raton, FL, 2015.

\bibitem{abba2006}
Abbasbandy, S. Homotopy perturbation method for quadratic Riccati differential equation and comparison with Adomian's decomposition method. \textit{Appl Meth Comput}. \textbf{2006}, \emph{17}, 485-490. doi:10.1016/j.amc.2005.02.014

\bibitem{khan2011}
Khan, N.A.; Ara, A.; Jamil, M. An efficient approach for solving the Riccati equation with fractional orders. \textit{Comput \& Math with Appl.} \textbf{2011}, \emph{61}, 2683-2680. doi: 10.1016/j.camwa.2011.03.017.

\bibitem{amin2010}
Aminkhah, H.; Hemmatenzhad, M. An efficient method for quadratic Riccatio differential equation. \textit{Commun Nonlinear Sci Numer Simul} \textbf{2010}, \emph{15}, 835-839. doi:10.1016/j.cnsns.2009.05.009.

\bibitem{li2010}
Li, Y.; Hu, L. Solving fractional Riccati differential equations using Haar wavelet. In: \textit{Third  International Conference on Information and Computing}. Wuxi, Jiang Su, China, June 4-6, 2010. doi: 10.1109/ICIC.2010.86.

\bibitem{svenkeson15} 
Svenkeson, A.; Glaz, B.; Stanton, S.; West, B.J. Spectral decomposition of nonlinear systems with memory. unpublished.

\bibitem{west15} 
West, B.J. Exact solution to fractional logistic equation. \textit{Physica A} \textbf{2015}, \emph{429}, 103-108. doi:10.1016/j.physa.2015.02.073.

\bibitem{zhang10} 
Zhang, S.; Zong, Q.A.; Liu, D.; Gao, Q. A generalized exp-function method for fractional Riccati differential equations. \textit{Commun Fract Calc.} \textbf{2010}, \emph{1}, 48.

\bibitem{west03}
West, B.J.; Bologna, M.; Grigolini, P. \textit{Physics of Fractal Operators}, 1st ed.; Springer: New York, 2003.

\bibitem{podlubny99}
Podlubny, I. \textit{Fractional Differential Equations}, 1st ed.; Academic: New York, 1999.

\bibitem{davis62}
Davis, H.T. \textit{Introduction to Nonlinear Differential and Integral Equations}, 1st ed.; Dover: New York, 1962.

\bibitem{jumarie06}
Jumarie, G. Modified Riemann-Liouville derivative and fractional Taylor series of nondifferentiable functions funther results. \textit{Comput \& Math with Appl}. \textbf{2006}, \emph{51}, 1367-1376. doi:10.1016/j.camwa.2006.02.001

\bibitem{kowalski91}
Kowalski, K.; Steeb, W.-H. \textit{Nonlinear Dynamical Systems and Carleman Linearization}, 1st ed.; World Scientific: Singapore, 1991.

\bibitem{diethelm2005}
Diethelm, K.; Ford, N.J.; Freed, A.D.; Luchko, Y. Algorithms for the fractional calculus: A selection of numerical methods. \textit{Comput Methods Appl Mech Engrg.} \textbf{2005}, \emph{194}, 743-773. doi: 10.1016/j.cma.2004.06.006.

\bibitem{diethelm2002}
Diethelm, K.; Ford, N.J.; Freed, A.D. A predictor-corrector approach for the numerical solution of fractional
differential equations. \textit{Nonlinear Dynam.} \textbf{2002}, \emph{29}, 3-22. doi: 10.1023/A:1016592219341.

\end{thebibliography}
\makeatletter
\renewcommand\@biblabel[1]{#1. }
\makeatother

\end{document}